\documentstyle[12pt]{article}
\input psfig
\begin{document}
\title{Information Theory, Quark Clusters in Nuclei, and Parton 
       Distributions}
\author{Athanasios N. Petridis \\
{\em Dept. of Physics and Astronomy, Iowa State University,} \\
{\em Ames, IA 50011}}

\maketitle

%%%%%%%%%%%%%%%%%%%%%%%%%%%%%%%%%%%%%%%%%%%%%%%%%%%%%%%%%%%%%%%%%%%%%%%%
%
% Abstract, PACS numbers and keywords
%
%%%%%%%%%%%%%%%%%%%%%%%%%%%%%%%%%%%%%%%%%%%%%%%%%%%%%%%%%%%%%%%%%%%%%%%%
\begin{abstract}

The procedure of maximization of information entropy can be used to
improve our knowledge of parton distributions. This method has been
applied in order to achieve improved description of the nuclear effect
in $\Upsilon$ production due to gluon distribution modification in
nuclei.
  
\vspace{0.20in}
\hspace{-0.28in}
PACS numbers: 24.85.+p, 13.85.Ni, 25.40.Ep, 25.75.-q \\
Keywords: information, quarkonium, suppression, shadowing

\end{abstract}

%%%%%%%%%%%%%%%%%%%%%%%%%%%%%%%%%%%%%%%%%%%%%%%%%%%%%%%%%%%%%%%%%%%%%%%%
%
% Main body
%
%%%%%%%%%%%%%%%%%%%%%%%%%%%%%%%%%%%%%%%%%%%%%%%%%%%%%%%%%%%%%%%%%%%%%%%%
\section{Introduction}

Information Theory was originally developed by Hartley~\cite{Hartley},
Nyquist~\cite{Nyquist}, and Shannon~\cite{Shannon} in order to establish 
a mathematical description of telecommunications and to understand how
information may be lost upon transmission over noisy channels. Shannon,
in particular, developed a complete formalism in which the concept of
information is quantified and important theorems regarding its transmission
are proven. The fundamental quantity that measures information is 
{\em information entropy}. Shannon has shown that the information
entropy is the most suitable function of the probabilities for emission of
signals by an ergodic source that measures the magnitude of the
receiver's uncertainty on those signals. In this sense information
entropy is a true measure of one's ignorance of their correct content.
The larger the entropy the greater the uncertainty and, consequently,
the information content. Therefore, maximizing the information 
entropy can lead to evaluation of the signal probability distribution
under the constraints imposed by the telecommunications problem at hand.

Application of this theory in physics could be versatile. All
quantum phenomena, for example, are stochastic in nature and are
described in terms of probability amplitudes. Then by appropriately
defining the information entropy of the physical system under 
consideration and maximizing it under constraints imposed by theoretical
assumptions or experimental data one can obtain the probability 
amplitudes that are most consistent with one's ignorance of the system.
This method has been used by Plastino~\cite{Plastino} to evaluate
wave functions for various physical systems. It is
a very powerful technique since it does not rely on any
specific modeling but only on what is actually known about the system
to derive best estimates for what is unknown.

In this article we apply Information Theory to improve parton, specifically
gluon, distributions in nuclei. These are suitable subjects to the method
because they are probabilistic in character. The problem whose solution we 
wish to improve is that of quarkonium suppression in proton-nucleus
collisions at very high energies. The production of charmonium states,
most notably of the $J/\Psi$ boson as well as of bottomonium ones,
especially the $\Upsilon$ resonances, has been observed in various
experiments involving heavy nuclear targets to be lower that in hydrogen
if the latter is multiplied by the mass number of the nucleus.
At energies achieved at Fermilab $J/\Psi$ and $\Upsilon$
suppression is very pronounced and exhibits a characteristic dependence
on the momentum fraction of the target nucleon carried by the struck
parton~\cite{Alde}. Many models have been developed to explain this
behavior. For the purposes of demonstrating the Information Theory
method we consider a model that is based on the assumption that quarks
in nuclei have a finite probability to conglomerate forming multi-quark
color singlet states, usually called (multi)quark clusters~\cite{Sato}.  
The parton distributions in such clusters differ from those in
single nucleons and generally are concentrated to lower momentum
fractions of the partons as the cluster becomes larger. This model 
supplemented by final state dissociation of the produced quarkonium
has successfully described $J/\Psi$ suppression in hadron-nucleus
collisions as well as most heavy ion collisions, the latter being
the topic of heated debate due to their relevance with Quark-Gluon
Plasma production. In this model very simple parton distributions
have been used based on very general assumptions. However, the gluon
distributions which play a crucial role in quarkonium production 
are poorly known in nuclei. It turns out that within this model the gluon 
distributions that solve the problem of $J/\Psi$ suppression are inadequate to
describe $\Upsilon$ suppression from the same experiment. We shall
use Information Theory to improve them in a manner that maintains
their applicability to the $J/\Psi$ data and, at the same time,
to enhance agreement with the $\Upsilon$ data. 

\section{Information Theory}

Suppose an ergodic source of information, which can be anything 
from a telegraphic device to a quantum system, produces signals
$x$ from some available ensemble $X$ with probability distribution
$p(x)$. We define the information entropy of the source as~\cite{Shannon}
\begin{equation}
S = - \sum_x p(x, \alpha_i) \; {\rm ln} \; p(x, \alpha_i),
\end{equation}
where the summation includes all instances of $x$ in $X$ and
can indicate an integral if $x$ is a continuous variable and
$\alpha_i$ are a group of fixed parameters in the function $p$. 
It is required that
\begin{equation}
\sum_x p(x, \alpha_i) = 1.
\end{equation}
If the logarithm is binary then $S$ is expressed in bits. Considered
as a functional of the probability distribution, $S$ is maximal
when $p$ is uniform; we know the most about a system when the
signals it produces have no variability. To obtain the optimal
function $p(x, \alpha_i)$, i.e., to estimate the ``best'' set of
parameters $\alpha_i$ we impose the extremization condition
\begin{equation}
\frac {\partial S}{\partial a_i} = 0, 
\end{equation}
for all $i$. To ensure a maximum second order derivatives must be 
looked at as well. In this work we assume a certain functional form for 
$p$ and simply want to determine its parameters. The procedure can be 
generalized to include cases in which we do not know the exact form of $p$ 
but there are constraints based on data. The method of Lagrange's 
multipliers can be used to evaluate distribution functions when a number 
of expectation values are known~\cite{Plastino}.

\section{Quarkonium Suppression in p-A Collisions}

\subsection{Facts and models}

The observed suppression of the $J/\Psi$, $\Psi^{\prime}$ production
cross section per unit mass number, $A$, in high energy 
hadron-nucleus~\cite{Alde,Kowitt} and nucleus-nucleus~\cite{Abreu} 
collisions exhibits a strong nuclear dependence. The hadron-nucleus
data also show that the depletion increases with the longitudinal momentum  
of the quarkonium. These results have generated many theoretical studies.
A variety of effects are thought to
contribute and numerous models have been suggested. These
contributions may be grouped into six major categories: (1) Quarkonium
scattering off parton and/or hadron co-movers~\cite{Mueller,Brodsky}; (2)
Glauber inelastic scattering on the nucleons~\cite{Claudie}; 
(3) Shadowing and EMC distortions of the nuclear parton
distributions~\cite{Gupta}; (4) Parton scattering before~\cite{Blaizot}
and after~\cite{Qiu} the hard process; (5) Parton energy loss in the initial 
and/or the final state~\cite{Sean} and, (6) Intrinsic charm in
nucleons~\cite{Brodsky}. 
Color transparency may alter the charmonium Glauber absorption~\cite{Farrar}.
In the case of heavy ion collisions contribution of an unconfined state 
(Quark Gluon Plasma) has probably been very small in earlier
experiments~\cite{Alde,Abreu} but is currently debated due to new 
data~\cite{NA50}. In our view more than one contribution must be carefully 
balanced to explain the quarkonium suppression. Here the modifications of 
the initial state parton distributions (for all values of the Bjorken 
variable, $x$) and the final state inelastic scattering (absorption)
are considered. 

In this article the term ``EMC effect'', named after the European Muon
Collaboration, will signify any deviation from unity of the structure
function ratio of a bound nucleon to that of a free one at any value of
the variable $x$, the latter defined as the fraction of the nucleon
momentum carried by the interacting parton.

\subsection{Quark clusters in nuclei}

The EMC effect has been studied in the framework of the expansion of a
nuclear state $|A\rangle$ on a complete basis of color singlet states
labeled by the number of (3, 6, 9 or more) valence quarks they contain.
Such states, also referred to as multi-quark clusters, are formed
when nucleons bound in a nucleus overlap sufficiently so that they share
their constituent partons. The probabilities for multi-quark cluster
formation can be estimated using nuclear wave functions~\cite{Sato} or
can be found by fitting DIS data~\cite{Havens,Lassila}. The agreement with
EMC and NMC~\cite{NMC} data is excellent down to $Q^2 \approx 2$ GeV$^2$
and the fit strongly constrains the quark momentum distributions in
clusters. (Momentum conservation in clusters fixes the {\em total\/}
momentum fraction carried by the gluons.) Good description of the small
enhancement above unity of the EMC ratio (antishadowing) around $x =
0.1$ requires inclusion of up to 12q clusters but the essential features
of the data are accommodated by a truncation to the 6q term and use of
an {\em effective\/} 6q cluster probability, $f$. The observed shadowing
of the structure function, $F_2$, in the nucleus for low $x_B$ and the
depletion for intermediate $x$ combined with the QCD sum rules require
the presence of antishadowing of $F_2$ for other values of $x$. This
antishadowing, however, need not be restricted to the range $0.05 < x
< 0.2$. In fact, this model predicts antishadowing for $x > 0.8$ in
agreement with the data of Ref.~\cite{NMC} on the slope of the Ca over
deuterium structure function ratio. In addition, the electron DIS
data from SLAC confirm the excellent agreement of this model with the
observed nuclear dependence for all $x > 0.1$~\cite{Gomez}. The model
has been successfully applied to explain nuclear effects in Drell-Yan
processes~\cite{DY} and gives interesting predictions for these effects
at RHIC energies~\cite{RHIC}. Direct photon production in hadron-nucleus
collisions~\cite{Dir} has also been predicted to be altered by nuclear
effects.

The probability $f$ depends on $A$ approximately logarithmically 
and is 0.040-0.052 for deuterium~\cite{Dir}. Very dense nuclei 
($^4$He) have $f$ values larger than the logarithmic prediction. 
In the scaling limit the Nq proton-like cluster parton momentum
distributions are assumed to have standard forms,
\begin{eqnarray}
V^{u,d}_N(x) &=& B^{u,d}_N \sqrt{x}(1-x)^{b^{u,d}_N},
                             \:\:\:   b^{d}_N = b^{u}_N+1, \\
S_N(x)       &=& A_N(1-x)^{a_N},\:\:\:\: A_N = x_N^S(1+a_N), \\
G_N(x)       &=& C_N(1-x)^{c_N},\:\:\:\: C_N = x_N^G(1+c_N).  
\end{eqnarray}
The exponents that best describe the NMC data are 
($b_3^u, a_3, b_6^u, a_6$) = (3, 9, 10, 11). 
For the gluon exponents the direct $\gamma$ data suggest
($c_3, c_6$) = (6, 10)~\cite{Dir}. The valence exponents approximately
follow the dimensional counting rules. The ocean consists of three
species of quarks (and their antiparticles) with the strange
distribution being half as large as the up (or down) quark sea
distribution. The gluon momentum fraction is taken 5 times larger than
the ocean one.  Isospin invariance relations connect the distributions
that belong to the same isospin states reducing, thus, the number of
independent paramenters. Kinematics forces the fraction of the cluster
momentum carried by a parton in 6q clusters, $x^{(6)}$,  to
be half as large as that in nucleons, $x^{(3)}$. In this model the QCD sum
rules are explicitly obeyed. Furhter description can be found
in Refs.~\cite{RHIC,Dir}. 

\subsection{Theoretical cross sections}

Including 3q and 6q clusters in the nucleus the doubly differential
hadron level cross section for quarkonium production
is calculated from the $\alpha_s^2$ order
parton level ones (with both quark annihilation ($q\bar{q}$)
and gluon fusion ($gg$) included).
The latter are convoluted with the appropriate sums of products of
parton distributions $H_{q\bar{q}}^{(i)}(x_1, x_2^{(i)})$ and
$H_{gg}^{(i)}(x_1, x_2^{(i)})$ ($i=3,6$), respectively~\cite{JPSI}.
The indices 1 and 2 refer to the probe (moving in the $+z$ direction in
the laboratory) and the target, respectively. The duality  
(in effect color evaporation) hypothesis is applied to integrate 
the differential cross section over $m^2$ with $m$ 
ranging from twice the $c (b)$ quark mass, $2m_{c(b)}$, to the open charm 
(bottom) threshold, $2m_{D(B)}$. Here $m_D = 1.864$ GeV and
$m_B = 5.278$ GeV. The duality constant, $F_d$, is simply the
portion of the total charmonium cross section (up to a given order)
that corresponds to the quarkonium; it cancels in cross section ratios. 
The resulting cross section can be expressed as a
function of the longitudinal momentum fraction carried by the produced
quark-antiquark pair, $x_F = 2p_L/\sqrt{s}$. The higher order corrections 
are assumed to result in a multiplicative factor, $K$. The transverse 
momentum dependence
due to higher order diagrams is, thus, integrated over and that due to
the intrinsic transverse momentum of the partons is neglected. These
considerations lead to the order $\alpha_s^2$ equation
\begin{equation}
\frac{d\sigma^{(A)}}{dx_F} = K \; F_d \; 
\int_{4m_{c(b)}^2}^{4m_{D(B)}^2} \; dm^2 \sum_{i=3,6} J^{(i)} \left [
H_{q\bar{q}}^{(i)} \; \hat{\sigma}_{q\bar{q}}(m^2) +
H_{gg}^{(i)} \; \hat{\sigma}_{gg}(m^2) \right ], 
\label{hadron}
\end{equation} 
where $\hat{\sigma}_{q\bar{q}}$ and $\hat{\sigma}_{gg}$ are the
partonic-level cross sections for the two hard processes and
$J^{(i)}$ are the Jacobians that transform $x_1$ and $x_2^{(i)}$ to
$x_F$ and $m^2$. In this way the entire quarkonium yield is found. The 
various final states being unitary rearrangements of one another have up 
to this point the same dependence on EMC-type nuclear effects at the
same $\sqrt{s}$. The perturbative $Q^2$ evolution of the parton 
distributions is omitted as it largely cancels in the calculation of
cross section ratios. In addition, the $Q^2$ values that are relevant to 
the production of quarkonium states are inside the scaling region
in which the parton distributions used in this work are valid.

\subsection{Final state absorption}

After its production the $c\bar{c}$ ($b\bar{b}$) system propagates in the 
nuclear medium developing into a quarkonium state which then decays into the
observed lepton pairs. During this stage the system may be inelastically
scattered by 3q and 6q clusters. It is reasonable 
to assume that the scattering occurs with clusters bound in the nucleus 
since the break up time of the latter exceeds the time the 
pair needs to traverse the nuclear radius. The $J/\Psi$ absorption cross
section has been measured to be $\sigma^{(3)}_{abs}\approx$ 3.5 
mbarn/nucleon~\cite{Anderson}. This number has been extracted in a
kinematic range in which parton distribution modifications are negligibly small
($x_2\approx 0.22$, the point at which the EMC ratio crosses the unit
line) and since it is the average absorption cross section over the path
traversed in the nucleus, color transparency effects are already in it.
The $\Psi^{\prime}$ is attenuated a little more than the
$J/\Psi$ due to its larger radius. In this It is also assumed
that the cross section for bottomonium ($\Upsilon$ states) absorption 
is of the same order as that of charmonium but smaller 
($\approx 3$ mbarn/nucleon) due to its more compact size.

The cross section on a 6q cluster is $2^{3/2}$ times larger than that on
a nucleon (bag model estimate) but the density of scattering centers in
the medium is reduced when $f\neq 0$. Then 
\begin{equation}
\langle \rho\sigma_{abs} \rangle 
= \sigma^{(3)}_{abs}\rho_A [(1-f)+2^{3/2}f]/(1+f), 
\end{equation}
where $\rho_A$ is the number density of the nucleus $A$ taken as
constant within the nuclear volume. The average path length the
$c\bar{c}$ ($b\bar{b}$) pair travels in an approximately spherical nucleus 
of radius $r_A$ estimated by means of a simple geometrical (not eikonal)
calculation is $L_A\approx 2r_A/\pi$.  The experimentally measured
nuclear RMS radii~\cite{Jagger} are employed to compute $\rho_A$ and
$L_A$. The cross section in Eq.(\ref{hadron}) is then attenuated by the
factor 
\begin{equation}
P_A = \exp[-\langle \rho \sigma_{abs}
\rangle L_A]. 
\end{equation}

\section{Charmonium Suppression}

The nuclear dependence is extracted by taking the ratio, $R_A$, of the
$J/\Psi$ production cross section per unit $A$ in collisions of a
hadron, proton in this case, with a heavy nucleus to that in collisions
with a light one and examining its $x_F$ dependence. Since the
$K$-factor may also depend on $x_F$ at first we examine
the ratio of the experimental to the order $\alpha_s^2$ theoretical
cross section. In Fig.~1 we present this ratio for Cu and Be using a
representative set of parameters for our model and the data of
Ref.~\cite{Kowitt}. $F_d$ is the same in both cases and cancels in the ratio.
Clearly the $K$-factor exhibits a strong $x_F$
dependence but, most importantly for our purposes, is the same for the
two nuclei. This implies that it will cancel in ratios of two
theoretical or experimental cross sections. The origin of this factor
must, thus, lie in processes that do not depend on the size and
intrinsic properties of the nucleus (the absorption part is included in
the results of Fig.~1). The $x_F$ dependence of $K$ in Fig.~1 agrees with 
that of the ratio of the ``diffractive'' to hard cross sections in
Ref.~\cite{Badier}. Issues related to the relative suppression 
of various charmonium states are outside the scope of this article.

The results for $R_A$ are confronted with the data of Ref.~\cite{Alde} in
Fig.~2. The dotted lines in Fig.~2 represent the prediction of the full
model including 6q clusters and final state absorption with the lowest
(highest) value of $f$ for the heavy (light) nucleus and the largest values
of the 6q ocean and gluon exponents, ($a_6, c_6$) = (12, 11); the solid
ones correspond to the opposite $f$ combination and ($a_6, c_6$) = (10,
9). In order to make the influence of the initial and final state
contributions to $R_A$ clear in Fig.~2(d) the results without final state
absorption are shown (short and long dash curves with the same
connotation as the dotted and solid ones, respectively) as well as the
result with only final state absorption, $f=0$ (dot-dash line). It is
evident that the predictions of the full model are in agreement
with the data. We note in passing that for large negative 
$x_F$ this model predicts antishadowing of $J/\Psi$ production in p-A 
collisions because in this region large $x_2$ values for the gluon
distribution ratio are accessed.

\section{Bottomonium Suppression}

Using the model we described earlier we can also calculate the suppression
ratio for the $\Upsilon$ states and compare the results with the
data of Ref.~\cite{Upsilon}. Specifically we calculate the exponent $\alpha$ 
defined by means of the equation
\begin{equation}
\frac{d \sigma^{(A)}}{d x_2} = A^{\alpha} \frac{d \sigma^{(d)}}{d x_2},
\end{equation}
where the superscripts refer to large nuclei ($A$) or deuterium ($d$) targets
and $x_2 = x_2^{(3)}$. At this point it is instructive to observe that gluon
fusion dominates the charmonium production process and is very important for
bottomonium production as well. Therefore, the
ratio of cross sections to a large extent reflects the ratio of gluon 
distributions, $R_G^{(A)}$. It is not hard to see that with the
given definition of gluon distributions $R_G^{(A)}$ monotonically increases
with $x_2$. As shown in Fig.~3, however, the data of Ref.~\cite{Upsilon}
contradict this theoretical prediction. At $x_2 \approx 0.15$ there is
a wide ``bump'' and the ratio starts decreasing at higher $x_2$. The reason
that this behavior becomes more apparent in the case of $\Upsilon$ production
is the fact that the $\Upsilon$ is much more massive than the $J/\Psi$.
For given center of mass energy, a particular quarkonium momentum, $x_F$,
probes a larger $x_2$ value in the $\Upsilon$ case. The gluon distributions
being the most relevant and the least known among all the partons should
be the first candidates for improvement.

\subsection{Improved gluon distributions}

At this point we turn to Information Theory. The total momentum fractions
carried by the partons in an Nq cluster, 
\begin{equation}
z^{(a)}_N = \int_0^1 dx F^{(a)}_N (x),
\end{equation}
where $a$ designates the type of parton and $F_N^{(a)}$ is the momentum 
distribution of $a$, must always add up to unity. Consequently the functions 
$F_N^{(a)}$ satisfy the condition required in order to define an information 
entropy for them, 
\begin{equation}
S_N =  -\sum_a \: \int^{1}_{0} \: dx \: F_N^{(a)} (x) \ln {F_N^{(a)}(x)}.
\end{equation}
 
We shall maintain the quark distributions as defined in the previous section
and modify the gluon distributions under the constraint that the total
momentum fraction carried by gluons in each type of cluster is fixed and
equal to that of the unmodified distributions, i.e., 
$z_3^{(g)} = 0.57$ for nucleons and $z_6^{(g)} = 0.60$ for 6q clusters.

The trend of the $\Upsilon$ data suggests that the simplest possible
modification to the gluon distributions is an alteration of the linear
term in $x$. We will, then, define corrected momentum distributions
for the gluons in nucleons and 6q clusters as
\begin{equation}
G_N(x) = C_N(1-x)^{c_N} + C_N^{\prime} x. 
\end{equation} 
For each $N$ we now have two unknown parameters $C_N$ and $C_N^{\prime}$.
Momentum conservation, i.e., that the fractions $z_N^{(g)}$ are constant, 
imposes the condition
\begin{equation}
C^{\prime}_N = 2z_N^{(g)} - \frac{2 C_N}{c_N + 1},
\end{equation}
where the exponents $c_N$ are kept fixed, $c_3 = 6$ and
$c_6 = 10$. Then we evaluate $C_N$ from the requirement
\begin{equation}
\frac{\partial S_N}{\partial C_N} = 0.
\end{equation}

The maximization procedure yields the following numbers:
$(C_3,\:C_3^{\prime}) = (1.163,\:0.812)$ and  
$(C_6,\:C_6^{\prime}) = (1.411,\:0.987)$.
We can compare these numbers with the uncorrected ones:
$(C_3,\:C_3^{\prime})_{uncorr} = (4.130,\:0.0)$ and  
$(C_6,\:C_6^{\prime})_{uncorr} = (6.624,\:0.0)$. We have thus
changed the shape of the function without affecting its integral,
the total gluon momentum. The behavior of the new functions
differs from that of the old ones mostly in the large $x$ region 
for which on the other hand we have little experimental 
data. We note that the ocean distributions could not
accommodate such type of alteration because deeply inelastic
scattering imposes a constraint on the ratio of neutron to
proton structure functions, $F_2^{(n)}(x)/F_2^{(p)}(x) 
\rightarrow 1/4$ as $x \rightarrow 1$.

Using the corrected distributions we can recalculate the exponent
$\alpha$ and compare the results with the data. With a final state
absorption cross section of 3 mbarn/nucleon and the new distributions
the agreement with the data is considerably improved as shown in Fig.~3
in which the upper curve corresponds to the uncorrected
model, the lower curve to the corrected model using only
gluon fusion contributions and the middle curve to the  
corrected model including gluon fusion and quark annihilation.
It must be pointed out that this is not the only model that
gives reasonable description of the relative $J/\Psi$ to $\Upsilon$ 
suppression data in p-A collisions. The authors of Ref.~\cite{Liuti}
attribute the smaller suppression of $\Upsilon$ to the
$Q^2 \approx m^2$ evolution of the distribution functions, where
$m$ is the mass of the resonance. Indeed, as discussed in Ref.~\cite{JPSI}
the evolution of the ocean (and consequently the gluon) distributions
leads to smaller shadowing as $Q^2$ decreases. Our model neglects 
the $Q^2$ evolution relying on the observation~\cite{JPSI} that
the masses of the quarkonium resonances are already in the scaling
region and attributes the reduced $\Upsilon$ suppression to its
smaller absorption cross section. It is conceivable that both
effects may, in fact, contribute to this phenomenon. We would mostly
like to emphasize that it is the shape of the suppression curve
that needed to be improved to account for the $\Upsilon$ data.  

It is worthwhile noting that due to the fact that $\Upsilon$
production probes a different kinematic regime from $J/\psi$
the correction on the gluon distributions has only a small 
effect on the $J/\Psi$ suppression curves. The agreement with 
the charmonium data is still good as it can be observed in Fig.~4,
although it slightly deteriorates at small $x_F$ (large $x_2$).
In addition the corrected model does not exhibit antishadowing 
of $J/\Psi$ production for negative $x_F$ a feature that would
be in contradiction with the data~\cite{Leitch}.  
Data on bottomonium and charmonium give complementary information 
on the gluon distributions in nuclei. In addition the $\Upsilon$ is a
really good probe of the initial state in which it is produced
due to low absorption cross section.

\subsection{Predictions for RHIC}

We can use this model to make predictions for the $J/\Psi$ and
$\Upsilon$ suppression at the Relativistic Heavy Ion Collider
(RHIC) with $\sqrt{s} = 200$ GeV/nucleon. The calculation
proceeds along the same lines as for p-A collisions but now
there is the additional possibility of 6q-6q cluster collisions.
The nuclear effect is, thus, more pronounced. The cross section
for collisions of a nucleus A with a nucleus B is~\cite{JPSI}
\begin{equation}
\frac{d\sigma^{(AB)}}{dx_F} = K F_d 
\int_{4m_c(b)^2}^{4m_{D(B)}^2} dm^2 \; \sum_{i=3,6} \sum_{j=3,6} J^{(i,j)} 
   \left [ H_{q\bar{q}}^{(i,j)} \; \hat{\sigma}_{q\bar{q}} +
H_{gg}^{(i,j)} \; \hat{\sigma}_{gg} \right ], 
\end{equation}
where $J^{(i,j)} = x_1^{(i)}x_2^{(j)}/[m^2(x_1^{(i)} + x_2^{(j)})]$,
$i$ and $j$ represent the type of colliding cluster, 
$\sqrt{s}$ is the nucleon-nucleon CM energy and $H^{(i,j)}_{q\bar{q}(gg)}$
are functions of parton distributions appropriate for A-B collisions.

In Fig.~5 we show theoretical results obtained with the uncorrected
(curves marked by (d))~\cite{JPSI} and the corrected (curves marked by (c))
models. The curves exhibiting less suppression are for the $\Upsilon$.
The corrected model leads to larger suppression. The large $x_F$ behavior 
is now much more distinct. It can be understood if we realize that
in symmetric heavy ion collisions as $x_F \rightarrow 1$ the large
$x$ region for the positively moving nucleus is probed. Therefore,
in the uncorrected model the ratio increases with $x_F$ reflecting
the increase in the gluon distribution ratio and exhibits
antishadowing while in the corrected one it flattens out and remains
below unity.

\section{Conclusions}

We have applied Information Theory to improve the gluon 
momentum distribution functions in nuclei, including the ``EMC effect''.
The main idea is that by defining an information entropy, $S$, for
those functions whose total integral is unity we can evaluate their
parameters by maximizing $S$ with respect to them. In other words
we assume that the best choice of parameters is the one that 
is consistent with maximal ignorance under the constraint of
momentum conservation. A quark-cluster model for the ``EMC effect''    
has been used to establish good agreement with the data on
$J/\Psi$ suppression in p-A collisions but proved inadequate to 
describe $\Upsilon$ suppression. Then Information Theory provided
us with a tool to improve the model with significant success.
The gluon distributions have been corrected for their behavior
at large $x$ and an overall agreement with the differential
cross sections for quarkonium suppression was achieved. 

An interesting aspect of this method is that it does not rely 
on any specific microscopic theory which in turn should be investigated
in detail but uses only very general notions. The solution
that is consistent with the assumption of maximal
ignorance, quantified by the information entropy, seems to be an optimal one.
This method can be used to improve the parameters of more detailed and
realistic parton distributions in nucleons and nuclei under constraints
imposed by experimental data. 

%%%%%%%%%%%%%%%%%%%%%%%%%%%%%%%%%%%%%%%%%%%%%%%%%%%%%%%%%%%%%%%%%%%%%%%%%%%%%
%
% Acknowledgements
%
%%%%%%%%%%%%%%%%%%%%%%%%%%%%%%%%%%%%%%%%%%%%%%%%%%%%%%%%%%%%%%%%%%%%%%%%%%%%%
The author would like to thank R. Vogt, S. Gavin and A. Plastino
for useful discussions.

%%%%%%%%%%%%%%%%%%%%%%%%%%%%%%%%%%%%%%%%%%%%%%%%%%%%%%%%%%%%%%%%%%%%%%%%%%%%%
%
% References
%
%%%%%%%%%%%%%%%%%%%%%%%%%%%%%%%%%%%%%%%%%%%%%%%%%%%%%%%%%%%%%%%%%%%%%%%%%%%%%
\bibliographystyle{unsrt}

%%%%%%%%%%%%%%%%%%%%%%%%%%%%%%%%%%%%%%%%%%%%%%%%%%%%%%%%%%%%%%%%%%%%%%%%%%%%%
%
% Figures and captions
%
%%%%%%%%%%%%%%%%%%%%%%%%%%%%%%%%%%%%%%%%%%%%%%%%%%%%%%%%%%%%%%%%%%%%%%%%%%%%%

% Fig. 1
\begin{figure}
\centerline{\psfig{figure=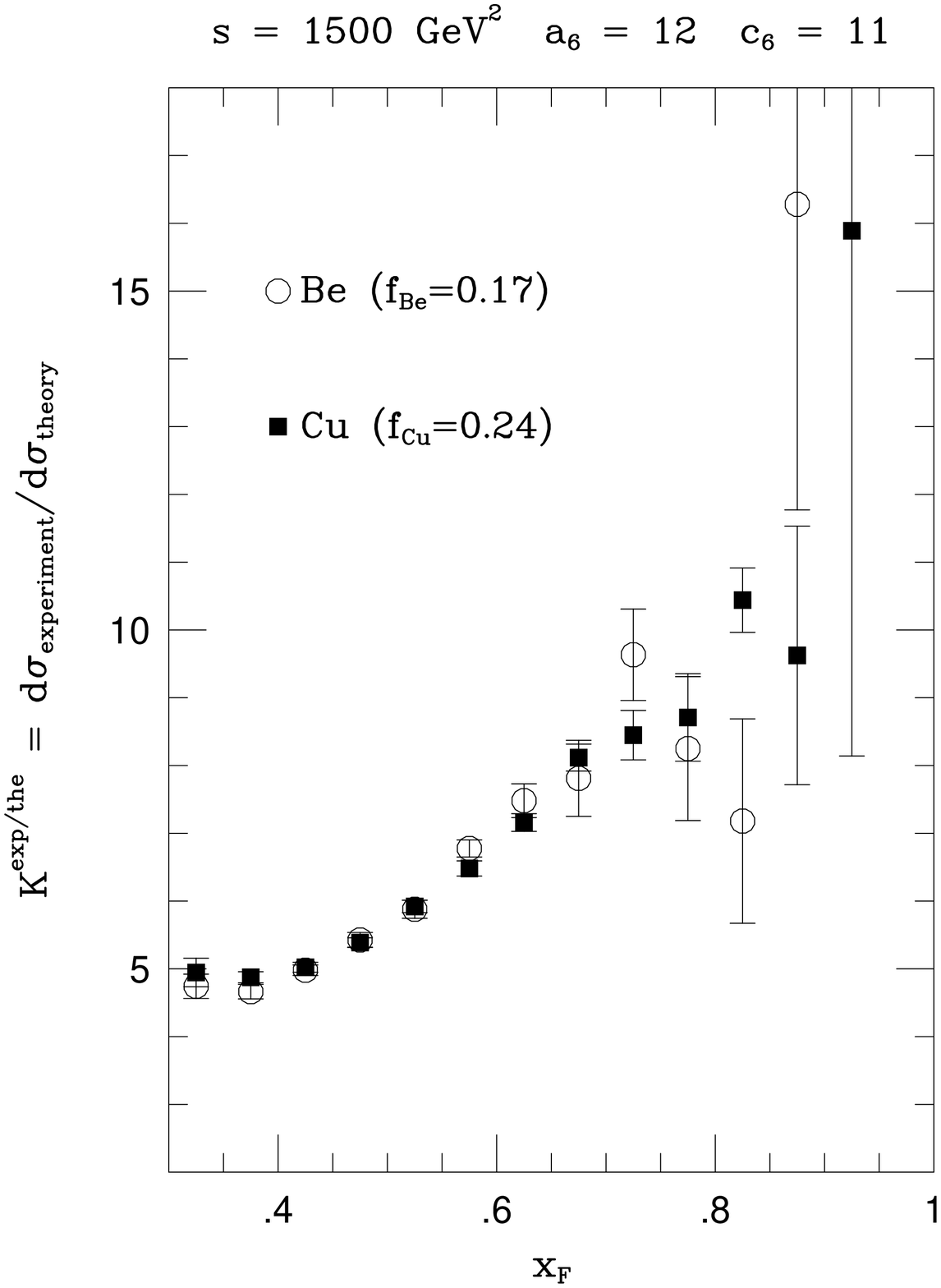,height=5.6in}}
\caption{The ratio of the experimental to the theoretical cross section
for $J/\Psi$ production in collisions of protons with Cu and Be nuclei.
The theoretical cross section includes cluster contributions and final
state absorption. The experimental cross section is from Ref.~\cite{Kowitt}
with statistical and systematic errors added in quadrature. The
uncertainty in the theoretical cross section is not included.}
\end{figure}

% Fig. 2
\begin{figure}
\vspace{-0.5in}
\centerline{\psfig{figure=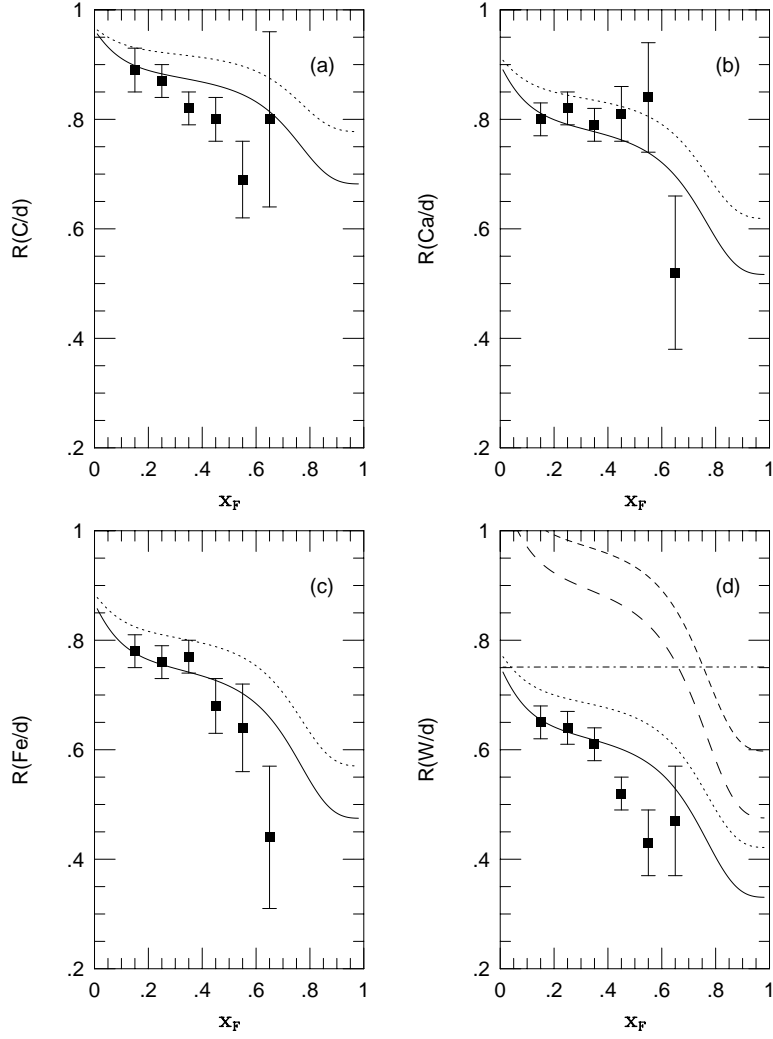,height=6.5in}}
\vspace{-0.16in}
\caption{C (a), Ca (b), Fe (c), and W (d) to deuterium $J/\Psi$ production 
ratio versus $x_F$. The 6q probabilities are evaluated using 
$f = k\ln A$ with 0.040-0.052 for deuterium. The 6q ocean and gluon exponents
are 10, 9, respectively for the solid lines and 12, 11 for the dotted ones.
Panel (d) also shows the results for no absorption (short and long dash 
curves) and with absorption only (dot-dash line). The data are from 
Ref.~\cite{Alde} at $\sqrt{s} = 38.7$ GeV. The errors are statistical.}
\end{figure}

% Fig. 3
\begin{figure}
\centerline{\psfig{figure=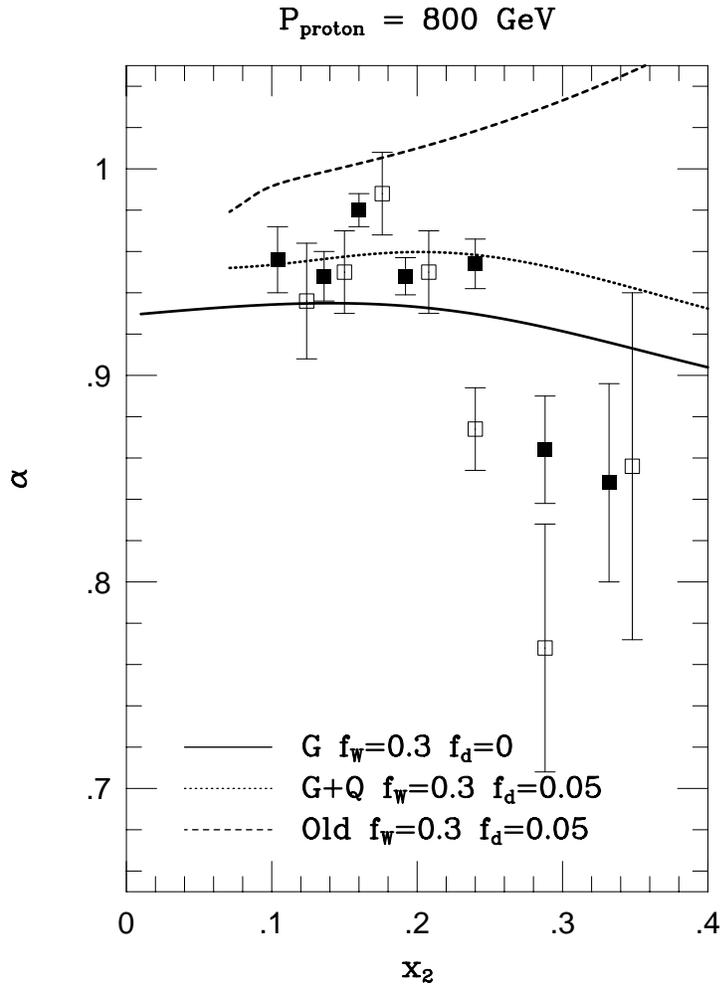,height=5.6in}}
\caption{Exponent of the W to deuterium $\Upsilon$ production ratio versus
$x_2$. The chosen 6q probabilities are shown in the plot. The data points are
from Ref.~\cite{Upsilon}. The solid squares are for the $1S$ state and the open
ones for $2S + 3S$. The upper curve corresponds to the uncorrected
model. The lower curve corresponds to the corrected model using only
gluon fusion contributions. The middle curve corresponds to the full 
corrected model including quark annihilation.} 
\end{figure}

% Fig. 4
\begin{figure}
\centerline{\psfig{figure=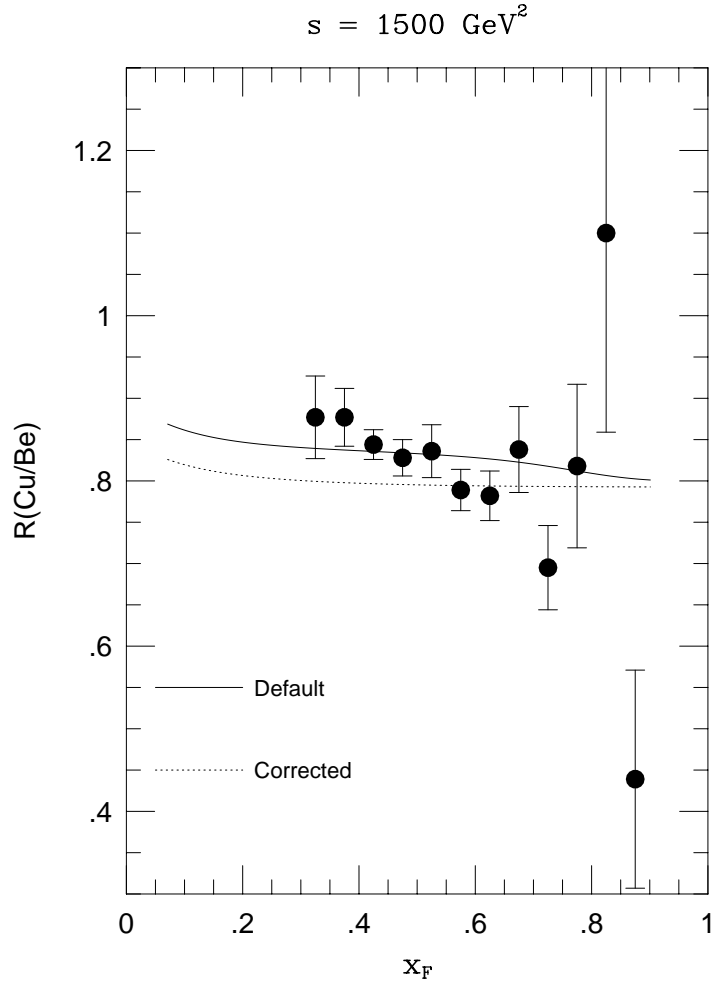,height=5.6in}}
\caption{The ratio of $J/\Psi$ production cross section per nucleon
on Cu to Be targets with a proton beam at $\sqrt{s} = 38.72$ GeV.
The solid curve is for the uncorrected gluon distributions
and the dash curve for the modified ones. The exponents used are
$(c_3, c_6)= (6,10)$ for the gluons and $(a_3, a_6) = (9,11)$ for the ocean.
The effective 6q cluster probabilities are 0.30 for Cu and
0.16 for Be. The data points are from Ref.~\cite{Kowitt}.}
\end{figure}

% Fig. 5
\begin{figure}
\centerline{\psfig{figure=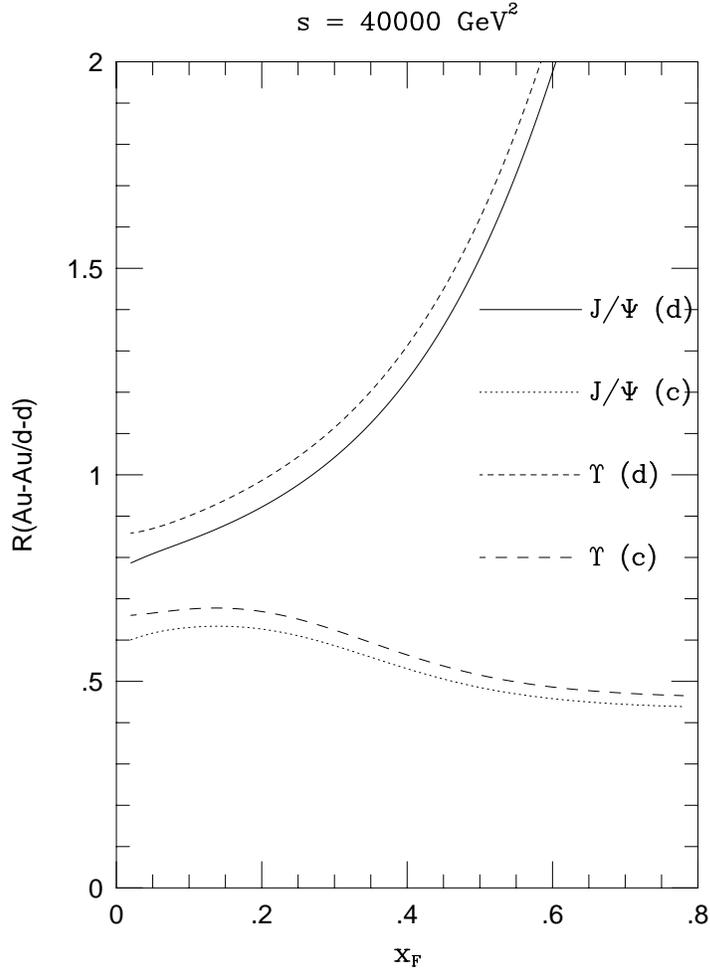,height=5.6in}}
\caption{The ratios of $J/\Psi$ and $\Upsilon$ production Au-Au cross sections
per nucleon over deuterium ones at RHIC energies. The curves marked by (d) 
are for the uncorrected model and those marked by (c) for the corrected one.
The effective 6q probabilities are 0.40 for Au and 0.05 for deuterium.
The used exponents are $(c_3, c_6)= (6,10)$ for the gluons and 
$(a_3, a_6) = (9,11)$ for the ocean. The absorption cross sections are 
3.5 mbarn/nucleon for the $J/\Psi$ and 3.0 mbarn/nucleon for the $\Upsilon$.}  
\end{figure}

\end{document}